\documentclass[notitlepage]{revtex4-1}
\usepackage{amsmath}
\usepackage{amssymb}
\usepackage{bm}
\usepackage[mathscr]{eucal}
\usepackage{theorem}
\usepackage{graphicx}
\usepackage{psfrag}
\usepackage{subfigure}
\usepackage{color}
\usepackage{wasysym}
\include{graphix}
\usepackage{framed}
\usepackage{enumitem}%
\usepackage{lipsum}
\usepackage{float}
\usepackage{MnSymbol}
\usepackage{url}

\graphicspath{{}}

\newcommand{\vecx}{\mathbf{x}}

\newcommand{\vecrhs}{\mathbf{f}}
\newcommand{\vecf}{\mathbf{f}}
\newcommand{\vecg}{\mathbf{g}}

\newcommand{\vecfhat}{\widehat{\mathbf{f}}}
\newcommand{\vech}{\mathbf{h}}
\newcommand{\veck}{\mathbf{k}}

\newcommand{\mxA}{\mathbf{A}}
\newcommand{\mxE}{\mathbf{E}}
\newcommand{\mxXhat}{\widehat{\mathbf{X}}}

\newcommand{\mxL}{\mathbf{L}}
\newcommand{\mxR}{\mathbf{R}}

\newcommand{\mathd}{\mathrm{d}}
\newcommand{\mathe}{\mathrm{e}}

\newtheorem{remark}{Remark}

\begin{document}
\title{cuPentBatch -- A batched pentadiagonal solver for NVIDIA GPUs}
\author{Andrew Gloster}
\affiliation{School of Mathematics and Statistics, University  College Dublin, Belfield, Dublin 4, Ireland}

\author{Lennon \'O N\'araigh}
\affiliation{School of Mathematics and Statistics, University  College Dublin, Belfield, Dublin 4, Ireland}
\author{Khang Ee Pang}
\affiliation{School of Mathematics and Statistics, University  College Dublin, Belfield, Dublin 4, Ireland}


\date{\today}

\begin{abstract}
We introduce cuPentBatch -- our own pentadiagonal solver for NVIDIA GPUs.  The development of cuPentBatch has been motivated by applications involving numerical solutions of parabolic partial differential equations, which we describe.  Our solver is written with batch processing in mind (as necessitated by parameter studies of various physical models).  In particular, our solver is directed at those problems where only the right-hand side of the matrix changes as the batch solutions are generated.  As such, we demonstrate that cuPentBatch outperforms the NVIDIA standard pentadiagonal batch solver gpsvInterleavedBatch for the class of physically-relevant computational problems encountered herein.

\vspace{0.1in}
\noindent{\bf{Program Summary}} \\
Program Title: cuPentBatch \url{https://github.com/munstermonster/cuPentBatch} \\
Licensing Provision: Apache License 2.0 \\
Programming Languages: C, C++, CUDA \\
Computer: Variable, equipped with CUDA capable GPU \\
Operating System: Linux, Mac and Windows \
\end{abstract}

\maketitle

\noindent Keywords: CUDA, Pentadiagonal, Hyperdiffusion, Cahn-Hilliard Equation, C, C++, Matrix Inversion
\vspace{0.1in}

\noindent \textbf{Minor edits shown in bold.}

\section{Introduction}
\label{sec:intro}
Graphics Processing Units (GPUs) are a hardware solution for a variety of parallel computing problems in Applied Mathematics and Computational Physics.  Due to the efficient way in which many programming tasks may be parallelised with GPUs, their use is growing.  This article considers the inversion of pentadiagonal linear systems using GPU computing.   We denote the generic problem here as $\mxA\vecx=\vecrhs$, where $\mxA$ is an $N\times N$ real invertible pentadiagonal matrix, $\vecrhs$ is a column vector of length $N$, and $\vecx$ is a column vector with  entries to be determined (we use $\vecrhs$ for the right-hand side, to be consistent with specific notation to be developed below).  However, as we are motivated by physical applications, we focus on a particular pentadiagonal linear problem, which arises in the context of numerical solutions of fourth-order (generalised) parabolic partial differential equations (PDEs) via finite-difference methods.  At the same time, we emphasise that the algorithms we develop are generic and carry over to arbitrary pentadiagonal systems.

The present work is based on the CUDA parallel computing platform.  As such, our aim herein is to implement the various finite-difference codes outlined below in Section~\ref{sec:model} with a CUDA code executed on an NVIDIA Titan X Pascal graphics card (with details provided below in Section~\ref{sec:implementation}).  For the purpose of performing a parametric study on the various finite-difference models, it is necessary to solve a large number of independent pentadiagonal systems -- essentially, solving the independent linear systems in batch mode.  This is a computationally-intensive task, and justifies the deployment of GPU computing.

CUDA already contains a library for solving pentadiagonal problems in batch mode, and the current state of the art algorithm is gpsvInterleavedBatch, which comes as part of the cuSPARSE library in CUDA.  The problem to be solved may be written in abstract terms as $\mxA\vecx_i=\vecrhs_i$, where the index $i$ labels the various pentadiagonal problems to be solved.  The application we have in mind is a parametric study, in which the vector $\vecrhs$ may depend on a physical parameter (or parameters); hence, the index $i$ labels different values taken by the parameter.  As such, the vectors $\vecx_i$ and $\vecrhs_i$ change as the index $i$ changes, but the matrix $\mxA$ is the same in each case.  In this context, use of gpsvInterleavedBatch is not appropriate, as gpsvInterleavedBatch  updates the entries of $\mxA$ for each instance of the linear problem, which leads to superfluous memory access and unnecessary computational overhead.  As such, the key point of the present work is to develop and test a new pentadiagonal batch 
solver (cuPentBatch) which leaves the matrix $\mxA$ intact for each instance of the pentadiagonal solver, thereby enhancing computational performance -- in short, we demonstrate that our newly developed cuPentBatch outperforms gpsvInterleavedBatch for the computational problems encountered herein.

The starting-point for developing the batched pentadiagonal solver is an existing batched tridiagonal solver called cuThomasBatch~\cite{cuThomasBatch}, based on the Thomas Algorithm, and now part of the CUDA library as gtsvInterleavedBatch.  We herein extend cuThomasBatch to accommodate pentadiagonal problems.  We also provide several examples from Computational Physics and Applied Mathematics where pentadiagonal problems naturally arise -- not only as computational problems to be solved on a one-off basis, but in the context of parametric studies, where solutions in batch mode are essential.  The pentadiagonal problems we exemplify are \textbf{symmetric positive definite} -- this justifies the use of an extended Thomas algorithm, which is numerically stable in precisely this setting.
%
%
%
%
%
The paper is organised as follows.  In Section~\ref{sec:model} we outline the numerical PDE-based models that provide the motivation for the development of the pentadiagonal solver.  We describe the pentadiagonal system to be solved and outline the algorithm for its solution.  We also present some sample numerical results with validation.  We introduce the parallel pentadiagonal solver in Section~\ref{sec:perf} and present the results of a  performance analysis --  we show how the present algorithm has superior performance to the existing in-house CUDA library (gpsvInterleavedBatch).  Concluding remarks are presented in Section~\ref{sec:conc}.

\section{Physical and Computational model}
\label{sec:model}

As we are motivated by key physical problems from  Applied Mathematics and Computational Physics, in this section we develop the new pentadiagonal solver in the context of physical models, namely the Cahn--Hilliard equation and the hyperdiffusion equation.  At the same time, we emphasise that the algorithms developed herein are generic and carry over to arbitrary pentadiagonal systems.

\subsection{The Cahn--Hilliard equation}
The Cahn--Hilliard equation models phase separation in a binary liquid: 
when a binary fluid in which both components are initially well mixed undergoes rapid cooling below a critical temperature, both phases spontaneously separate to form domains rich in the fluid's component parts~\cite{CH_orig}. The domains expand over time in a phenomenon known as coarsening~\cite{Bray_advphys}.   The equation is used as a model in polymer physics~\cite{Hashimoto} and interfacial flows~\cite{ding2007diffuse}.  We present the mathematical framework for the Cahn--Hilliard equation in what follows.  For definiteness, we work in one spatial dimension, although the theory readily carries over to multiple spatial dimensions (we briefly outline the computational methodology necessary for work in higher dimensions in Section~\ref{sec:conc} below).  As such, a single scalar concentration field $C(x,t)$ characterises the binary mixture, and a concentration level $C=\pm 1$ indicates phase separation of the mixture into one or other of its component parts, while $C=0$ denotes a perfectly mixed state.  The free energy for the mixture can be modelled as $F[C]=\int_\Omega \left[(1/4)(C^2 -1)^2 +(1/2)\gamma(\partial C/\partial x)^2\right] \mathd x$, where the first term promotes demixing and the second term smooths out sharp gradients in transition zones between demixed regions; also, $\gamma$ is a positive constant, $\Omega$ is the container where the binary fluid resides.  The twin constraints of mass conservation and energy minimisation suggest a gradient-flow dynamics for the evolution of the concentration: 
$\partial_t C = \partial_x\left[D(C)\partial_x(\delta F/\delta C)\right]$, where $\delta F/\delta C$ denotes the functional derivative of the free energy and $D(C)\geq 0$ is the mobility function, assumed for simplicity in this work to be a positive constant.  As such, the basic model equation reads
\begin{subequations}
\begin{equation}
\frac{\partial C}{\partial t}=D\frac{\partial^2}{\partial x^2}\left(C^3-C-\gamma\frac{\partial^2 C}{\partial x^2}\right),\qquad x\in \Omega,\qquad t>0.
\label{eq:ch_basic}
\end{equation}
The initial condition is given as
\begin{equation}
C(x,t=0)=f(x),\qquad x\in \overline{\Omega}.
\end{equation}%
\label{eq:ch_all}%
\end{subequations}%
For simplicity, we focus on the case where $\Omega=(0,L)$, with periodic boundary conditions, such that $C(x+L,t)=C(x,t)$, for all $t>0$ and $x\in\Omega$.

A well-established simulation method in this scenario is to discretise Equation~\eqref{eq:ch_basic} in time using a semi-implicit scheme as follows~\cite{Zhu_numerics,naraigh2007bubbles}:
\begin{equation}
\frac{C^{n+1}-C^n}{\Delta t}=\left[D\frac{\partial^2}{\partial x^2}\left(C^3-C\right)\right]^n-\gamma D\frac{\partial^4 C^{n+1}}{\partial x^4},
\label{eq:temporal}
\end{equation}
where the superscript $n$ indicates evaluation at time $t=n\Delta t$, where $\Delta t$ is the time step.  As such,
\[
C^n=C(x,t=n\Delta t),\qquad C^{n+1}=C(x,t=(n+1)\Delta t),
\]
and similarly for the term $\left[D\partial_{xx} \cdots\right]^n$ in Equation~\eqref{eq:temporal}.
Equation~\eqref{eq:ch_all} and its discretised version~\eqref{eq:temporal} are of great relevance for mathematical modelling, and are non-trivial to solve.  However, for the purposes of numerical methods, it is of interest for the present to focus on the hyperdiffusion equation, which can be obtained by setting the phase-separation term $C^3-C$ to zero in Equation~\eqref{eq:ch_all}.  This provides a convenient and simple basis from which to start the discussion about the application of pentadiagonal solvers to mathematical models.   A second compelling reason for studying the simplified (linear) hyperdiffusion equation is that it possesses explicit analytical solutions, which can be used as a rigorous test of our numerical methods.  We outline this approach in what follows below.

\begin{remark}
The Cahn--Hilliard equation~\eqref{eq:ch_all} with forcing is used as a model of phase separation in the presence of temperature gradients~\cite{weith2009traveling}, and reads as follows:
\begin{equation}
\frac{\partial C}{\partial t}=D\frac{\partial^2}{\partial x^2}\left(C^3-C-\gamma\frac{\partial^2 C}{\partial x^2}\right)+\Phi(x,t,\mu_1,\mu_2,\cdots),
\label{eq:forcing}
\end{equation}
where $\Phi$ denotes the forcing function and $\mu_1$, $\mu_2$ etc. are parameters.  A parameter study based on the model~\eqref{eq:forcing} provides the motivation for batch programming in this work.  As such, different solutions of Equation~\eqref{eq:forcing} corresponding to different parameter values $(\mu_1,\mu_2,\cdots)$ can be computed in parallel in batch mode, with a view to understanding how the variation in the parameters affects the structure of the PDE solutions.
\end{remark}

\subsection{The hyperdiffusion equation}

Based on the motivation given above, we focus on the following hyperdiffusion equation in one spatial dimension:
\begin{equation}
\frac{\partial C}{\partial t}=-\gamma D \frac{\partial^4 C}{\partial x^4},\qquad t>0,\qquad x\in (0,L),
\label{eq:hyperdiff}
\end{equation}
with periodic boundary condition $C(x+L,t)=C(x)$ and initial condition $C(x,t=0)=f(x)$, valid on $[0,L]$.  We henceforth rescale the space and time variables; this is equivalent to setting $\gamma=D=L=1$.  We discrete Equation~\eqref{eq:hyperdiff} in space using centred differences and in time using the Crank--Nicolson method.  We use standard notation for the discretisation, with
\[
C_i^n = C(x=i\Delta x ,t=n\Delta t),
\]
where $\Delta x$ is the grid spacing in the $x$-direction.  The grid spacing, the problem domain length $L$ and the number of unknowns $N$ are related through $\Delta x=L/N$.    In this way, the discretised version of Equation~\eqref{eq:hyperdiff} is written as
\begin{multline}
\frac{C_i^{n+1}-C_i}{\Delta t}=
-\tfrac{1}{2}\Delta x^{-4}\left[C_{i+2}^{n+1}-4C_{i+1}^{n+1}+6C_{i}^{n+1}-4C_{i-1}^{n+1}+C_{i-2}^{n+1}\right]\\
-\tfrac{1}{2}\Delta x^{-4}\left[C_{i+2}^{n}  -4C_{i+1}^{n}  +6C_{i}^{n}  -4C_{i-1}^{n}+C_{i-2}^{n}\right].
\label{eq:disc}
\end{multline}
Upon rearranging terms, Equation~\eqref{eq:disc} can be written more compactly as follows:
\begin{multline}
\sigma_x C_{i - 2}^{n + 1} - 4 \sigma_x C_{i - 1}^{n+1} + (1 + 6 \sigma_x)C_{i}^{n+1} - 4 \sigma_x C_{i+1}^{n+1} + \sigma_x C_{i+2}^{n+1} \\ 
=  - \sigma_x C_{i - 2}^{n} + 4 \sigma_x C_{i - 1}^{n} + (1 - 6 \sigma_x)C_{i}^{n} + 4 \sigma_x C_{i+1}^{n} - \sigma_x C_{i+2}^{n},
\label{eq:1d_hyper_scheme}
\end{multline} 
where $\sigma_x = \Delta t / 2\Delta x^4$. 
\begin{remark}
With the Crank--Nicolson temporal discretisation and the centred spatial discretisation, the truncation error in the hyperdiffusion equation~\eqref{eq:1d_hyper_scheme} is $O(\Delta t^2, \Delta x^2)$. \textbf{It can also be shown that this discretisation is unconditionally stable using Von Neumann stability analysis.}
\end{remark}
\begin{remark}
A finite-difference approximation of the Heat Equation $\partial_t C=\partial_{xx} C$ with Crank--Nicolson temporal discretisation and high-order accurate spatial discretisation (specifically, involving nearest neighbours and next-nearest-neighbours on the spatial grid) also produces a pentadiagonal problem that can solved with the methods developed herein.
\end{remark}
We conclude this section by emphasising that both the Cahn--Hilliard and hyperdiffusion equations fall into the category of fourth-order parabolic PDEs~\cite{Elliott_Zheng}, as the highest-order derivative term appears in a a linear fashion (specifically, through the appearance of the operator $\mathcal{L}=-\gamma D\partial_{xxxx}$).    The linear operator $\mathcal{L}$ satisfies the generalised parabolic property
\[
\langle \phi ,\mathcal{L}\phi\rangle=\int_0^L \phi \left(\mathcal{L}\phi\right)\,\mathd x
=-\gamma D\int_0^L |\partial_{xx}\phi|^2\,\mathd x\leq 0,
\] 
i.e. $\langle \phi,\mathcal{L}\phi\rangle\leq 0$ for all non-zero smooth real-valued $L$-periodic functions $\phi(x)$.

\subsection{The pentadiagonal matrix system}

Equation~\eqref{eq:1d_hyper_scheme} can be rewritten as a pentadiagonal matrix system, modulo some off-diagonal terms to deal with the periodic boundary conditions:
\begin{subequations}
\begin{equation}
\underbrace{
\begin{pmatrix}
c & d & e & 0 & \cdots &  0 & a & b \\
b & c & d & e & 0 & \cdots &   0 & a \\
a & b & c & d & e & 0 & \cdots  & 0\\
0 & \ddots & \ddots & \ddots & \ddots & \ddots & \ddots &   \vdots\\
\vdots& \ddots & \ddots & \ddots & \ddots & \ddots & \ddots & 0 \\
0 & \cdots  & 0  &  a  & b  & c & d & e \\
e & 0 &  & 0   & a  & b & c & d \\
d & e & 0  &   \cdots & 0 &  a & b & c
\end{pmatrix}
}_{=\mxA}
\underbrace{
\begin{pmatrix}
x_{1} \\
x_{2} \\
\vdots \\
\vdots \\
\vdots \\
x_{N-2} \\
x_{N-1} \\
x_{N}
\end{pmatrix}
}_{=\vecx}
=
\underbrace{
\begin{pmatrix}
f_{1} \\
f_{2} \\
\vdots \\
\vdots \\
\vdots \\
f_{N-2} \\
f_{N-1} \\
f_{N}
\end{pmatrix}
}_{=\vecrhs}.
\label{eq:matrix_sys1}
\end{equation}
Here, the coefficients of the matrix in Equation~\eqref{eq:matrix_sys1} have the following meaning:
\begin{equation}
a  = \sigma_x,\qquad b = - 4 \sigma_x,\qquad
c = 1 + 6\sigma_x, \qquad
d= - 4 \sigma_x, \qquad e = \sigma_x.
\end{equation}%
Similarly,
\begin{equation}
f_i = - \sigma_x C_{i - 2}^{n} + 4 \sigma_x C_{i - 1}^{n} + (1 - 6 \sigma_x)C_{i}^{n} + 4 \sigma_x C_{i+1}^{n} - \sigma_x C_{i+2}^{n}.
\end{equation}%
\label{eq:matrix_sys}%
\end{subequations}%
As such, by inverting the matrix~\eqref{eq:matrix_sys}, the solution of the hyperdiffusion equation is advanced from time step $n$ to time step $n+1$.  Here, information concerning $C$ at time step $n$ is contained in the vector $\vecrhs$, 
from which $C$ at time step  $n+1$ is extracted via the vector $\vecx$.

The matrix~\eqref{eq:matrix_sys} can be inverted using any standard method but our focus is now on using a specific pentadiagonal solver~\cite{numalgC}. This though requires us to re-examine the matrix above which has terms that lie off the diagonal, thus an additional step will be required to remove these terms.  For this purpose, we use the algorithm of Navon in Reference~\cite{navon_pent}.   As such, the matrix $\mxA$ is decomposed such that the last two rows and last two columns are eliminated, this then reduces the matrix to a pure pentadiagonal form that can be solved along with an additional smaller solve to deal with the eliminated points in the matrix. 
Therefore, following the discussion in Reference~\cite{navon_pent} we introduce the matrix $\mxE$ which is simply an $(N - 2) \times (N-2)$ reduced version of $\mxA$, removing the last two rows and columns:
\begin{equation}
\mxE = 
\begin{pmatrix}
c & d & e & 0 & \cdots &  0 & \cdots & 0 \\
b & c & d & e & 0 & \cdots & \cdots   & \vdots \\
a & b & c & d & e & 0 & \cdots  & 0\\
0 & \ddots & \ddots & \ddots & \ddots & \ddots & \ddots &   \vdots\\
\vdots& \ddots & \ddots & \ddots & \ddots & \ddots & \ddots & 0 \\
0 & \cdots  & 0  &  a  & b  & c & d & e \\
0 & \cdots & \cdots  & 0   & a  & b & c & d \\
0 & \cdots & \cdots  &   \cdots & 0 &  a & b & c
\end{pmatrix}
\end{equation}
We define the following vectors based on these eliminations, all of row dimension $(N-2)$:
\begin{equation}
\mxXhat = 
\begin{pmatrix}
x_{1}^{n+1} \\
x_{2}^{n+1} \\
\vdots \\
\vdots \\
\vdots \\
x_{N-2}^{n+1}
\end{pmatrix},
\qquad
\vech =
\begin{pmatrix}
e & d \\
0 & e\\
0 & 0 \\
\vdots \\
a & 0 \\
b & a
\end{pmatrix},
\qquad
\vecfhat=
\begin{pmatrix}
f_{1} \\
f_{2} \\
\vdots \\
\vdots \\
\vdots \\
f_{N-2}
\end{pmatrix},
\qquad
\veck=
\begin{pmatrix}
a & b \\
0 & a\\
0 & 0 \\
\vdots \\
e & 0 \\
d & e
\end{pmatrix}.
\end{equation}
We have therefore reduced our system~\eqref{eq:matrix_sys} to two coupled simultaneous equations that can be written as follows:
\begin{subequations}
\begin{align}
\mxE\mxXhat + \veck 
\begin{pmatrix}
x_{N - 1} \\
x_{N}
\end{pmatrix}
= \vecfhat
\\
\vech^T \mxXhat + 
\begin{pmatrix}
c & d \\
b & c
\end{pmatrix}
\begin{pmatrix}
x_{N - 1} \\
x_{N}
\end{pmatrix}
=
\begin{pmatrix}
f_{N - 1} \\
f_{N}
\end{pmatrix}.
\label{eq:xhat}
\end{align}%
\end{subequations}%
We can solve for $\mxXhat$ in the first simultaneous equation through a pentadiagonal inversion of E and obtain:
\begin{equation}
\mxXhat = \mxE^{-1} \left[\vecfhat - \veck 
\begin{pmatrix}
x_{N - 1} \\
x_{N}
\end{pmatrix}
\right]
\label{eq:solve}
\end{equation}
Equation~\eqref{eq:solve} is substituted into Equation~\eqref{eq:xhat}.  After some rearrangement of terms, these operations yield an expression for the final two unknowns:
\begin{equation}
\begin{pmatrix}
x_{N - 1} \\
x_{N}
\end{pmatrix} = 
\left[
\begin{pmatrix}
c & d \\
b & c
\end{pmatrix} 
- \vech^T \mxE^{-1}\veck\right]^{-1}
\left[
\begin{pmatrix}
f_{N - 1} \\
f_{N}
\end{pmatrix}
- \vech^T\mxE^{-1}\vecfhat
\right].
\label{eq:first_two}
\end{equation}
As such, we solve for the final two unknowns first (via Equation~\eqref{eq:first_two}).  We then substitute the result for $(x_{N-1},x_N)^T$) into Equation~\eqref{eq:solve} and then invert to yield the entire vector $\vecx$.  Computationally the expressions for the inverted matrix in~\eqref{eq:first_two} and the $\vech^T\mxE^{-1}$ can be computed and stored at the start of any code to be reused as required. This eliminates much of the overhead for each time step of the hyperdiffusion algorithm.  In particular we make use of the fact that $(\mxE^{-1})^T\vech = \vech^T\mxE^{-1}$ to simplify this process further.

\subsection{Solution of the pentadiagonal system}
In this section we describe a standard numerical method~\cite{numalgC} for solving a pentadiagonal problem $\mxA\vecx=\vecrhs$.  We present the algorithm in a general context (in particular, independent of the earlier discussion on finite-difference solutions of PDEs).  As such,  in this section we assume that $\mxA$ is strictly pentadiagonal  with arbitrary nonzero entries, such that
 \begin{equation*}
\mxA = 
\begin{pmatrix}
c_1 & d_1 & e_1 & 0 & \cdots &  0 & \cdots & 0 \\
b_2 & c_2 & d_2 & e_2 & 0 & \cdots & \cdots   & \vdots \\
a_3 & b_3 & c_3 & d_3 & e_3 & 0 & \cdots  & 0\\
0 & \ddots & \ddots & \ddots & \ddots & \ddots & \ddots &   \vdots\\
\vdots& \ddots & \ddots & \ddots & \ddots & \ddots & \ddots & 0 \\
0 & \cdots  & 0  &  a_{N - 2}  & b_{N - 2}  & c_{N - 2} & d_{N - 2} & e_{N - 2} \\
0 & \cdots & \cdots  & 0   & a_{N - 1}  & b_{N - 1} & c_{N - 1} & d _{N - 1}\\
0 & \cdots & \cdots  &   \cdots & 0 &  a_{N} & b_{N}  & c_{N} 
\end{pmatrix}.
\end{equation*}
%
%
Three steps are required to solve the system:
\begin{enumerate}
\item Factor $\mxA = \mxL\mxR$  to obtain $\mxL$ and $\mxR$.
\item Find $\vecg$ from $\vecf = \mxL\vecg$
\item Back-substitute to find $\vecx$ from $\mxR\vecx = \vecg$
\end{enumerate}
Here, $\mxL$, $\mxR$ and $\vecg$ are given by the following equations:
\begin{subequations}
\begin{equation}
\mxL = 
\begin{pmatrix}
\alpha_1 &  &  &  &  &   &   \\
\beta_2 & \alpha_2 &  &  &  &  &     \\
\epsilon_3 & \beta_3 & \alpha_3 &  &  &  &   \\
 & \ddots & \ddots & \ddots &  &  &     \\
 &  &  \epsilon_{N - 1}  & \beta_{N - 1}  & \alpha_{N - 2} &    \\
&  &    & \epsilon_{N - 1}  & \beta_{N - 1} & \alpha_{N }\\
\end{pmatrix},
\qquad
\vecg = \begin{pmatrix}
g_{1} \\
g_{2} \\
\vdots \\
\vdots \\
g_{N-1} \\
g_{N}
\end{pmatrix},
\end{equation}
\begin{equation}
\mxR = 
\begin{pmatrix}
1 & \gamma_1  & \delta_1  &  &  &   &   \\
 & 1 & \gamma_2  & \delta_2  &  &  &     \\
 &  & \ddots & \ddots & \ddots  &  &   \\
 & & & 1  & \gamma_{N-2}  & \delta_{N-2}    \\
 &  &  &  & 1 & \gamma_{N-1}    \\
&  &    & &  & 1\\
\end{pmatrix}
\end{equation}%
\end{subequations}%
(the other entries in $\mxL$ and $\mxR$ are zero).  
The explicit factorisation steps for the factorisation $\mxA=\mxL\mxR$ are as follows:
\begin{enumerate}
\item $\alpha_1 = c_1$
\item $\gamma_1 = \frac{d_1}{\alpha_1}$
\item $\delta_1 = \frac{e_1}{\alpha_1}$
\item $\beta_2 = b_2$
\item $\alpha_2 = c_2 - \beta_2\gamma_1$
\item $\gamma_2 = \frac{d_2 - \beta_2 \delta_1}{\alpha_2}$
\item $\delta_2 = \frac{e_2}{\alpha_2}$
\item For each $i = 3, \cdots, N-2$
\begin{enumerate}[label*=\arabic*.]
\item $\beta_i = b_i - a_i \gamma_{i-2}$
\item $\alpha_i = c_i - a_i\delta_{i-2} - \beta_i \gamma_{i-1}$
\item $\gamma_i = \frac{d_i - \beta_i \delta_{i-1}}{\alpha_i}$
\item $\delta_i = \frac{e_i}{\alpha_i}$
\end{enumerate}
\item $\beta_{N-1} = b_{N-1} - a_{N - 1}\gamma_{N-3}$
\item $\alpha_{N - 1} =  c_{N-1} - a_{N-1}\delta_{N-3} - \beta_{N-1}\gamma_{N-2}$
\item $\gamma_{N-1} = \frac{d_{N-1}-\beta_{N-1}\delta_{N-2}}{\alpha_{N-1}}$
\item $\beta_{N} = b_{N} - a_{N }\gamma_{N-2}$
\item $\alpha_{N} =  c_{N}- a_{N}\delta_{N-2} - \beta_{N}\gamma_{N-1}$
\item $\epsilon_i = a_i, \quad \forall i$
\end{enumerate}
The steps to find $\vecg$ are as follows:
\begin{enumerate}
\item $g_1 = \frac{f_1}{ \alpha_1}$
\item $g_2 = \frac{f_2 - \beta_2 g_1}{\alpha_2}$
\item  $g_i = \frac{f_i - \epsilon_i g_{i-2} - \beta_i g_{i - 1}}{\alpha_i} \quad \forall i = 3 \cdots N$
\end{enumerate}
Finally, the back-substitution steps  find $\vecx$ are as follows:
\begin{enumerate}
\item $x_N = g_N$
\item $x_{N-1} = g_{N-1} - \gamma_{N-1}x_N$
\item  $x_i = g_i - \gamma_i x_{i+1} - \delta_{i}x_{i+2} \quad \forall i = (N-2) \cdots 1$
\end{enumerate}
In this work, we implement this algorithm in serial and parallel batch. It can be easily seen that only six vectors are required to implement this algorithm: five for the left-hand side and one for the right-hand side.  
In the initial factorisation step $\mxA=\mxL\mxR$ we overwrite the input matrix $\mxA$ with the factorised matrices $\mxL$ and $\mxR$ which can then be used for the inversion steps later, this is done to minimise memory usage.
%
%
%
It should be noted that this method is $O(N)$ and each system of equations in the batch must be solved serially by a thread. 

\subsection{Validation of Scheme}

\begin{figure}
	\centering
		\includegraphics[width=0.6\textwidth]{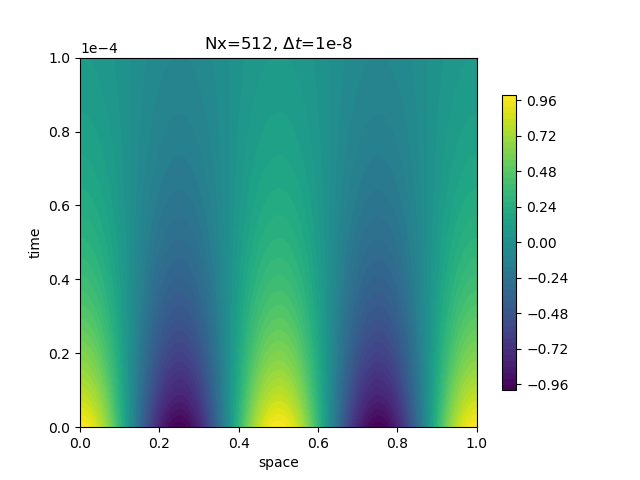}
		\caption{Space-time plot of the solution of the hyperdiffusion equation for a final time $T=10^{-4}$.  The model parameters and the initial condition are given in the main text.}
	\label{fig:1d_hyperdiffusion_sol}
\end{figure}
We have validated the implicit finite-difference method~\eqref{eq:disc}--\eqref{eq:1d_hyper_scheme} for the hyperdiffusion equation.  We use the pentadiagonal solver developed above.  As a first validation step, we have implemented the numerical algorithm in a serial C code.  This serves as a base case against which to compare the performance of the GPU code in what follows.
An advantage of performing validation tests with the hyperdiffusion equation is that the hyperdiffusion equation admits exact solutions.  As such, a harmonic initial condition $C(x,t=0)= A\cos(kx+\varphi)$ (with constant amplitude $A$, wavenumber $k=(2\pi/L)n$ and phase $\varphi$ evolves into an exponentially-damped harmonic solution for $t>0$, 
\begin{equation}
C(x,t)=A\mathe^{-\lambda t}\cos(kx+\varphi),\qquad t>0.
\label{eq:exact}
\end{equation}  
Here, $n$ is a positive integer, and $\lambda=\gamma D k^4$ is the known analytical decay rate.  In this section we work with $\gamma=D=L=1$.  We also take $A=1$, $\varphi=0$, and $n=2$. 

\begin{figure}
	\centering
		\includegraphics[width=0.6\textwidth]{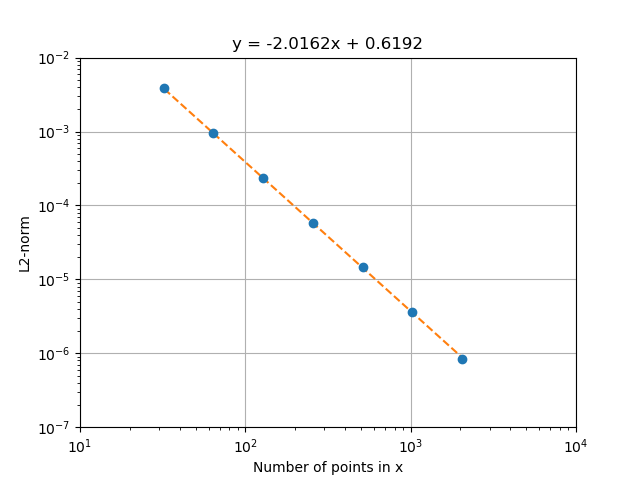}
		\caption{A plot of $\epsilon_N(T)$ as a function of $N$.  The final time is $T=10^{-4}$ and the time step is $\Delta t$ is $10^{-8}$.  The model parameters and the initial condition are given in the main text.  The line of best fit (on a log-log scale) is fitted and yields $\epsilon_N(T)\propto N^{-2.0162}$, compared to the theoretical value  $\epsilon_N(T)\propto N^{-2}$. }
	\label{fig:1d_hyperdiffusion_convergence}
\end{figure}

Based on this numerical setup, a spacetime plot of the numerical solution $C(x,t)$ is shown in Figure~\ref{fig:1d_hyperdiffusion_sol}, starting at $t=0$, and ending at the final time $T=10^{-4}$.  The amplitude numerical solution exhibits a rapid decay in time, consistent with the exact solution~\eqref{eq:exact}. 
We further examine the $L^2$ norm of the absolute error $\epsilon_N(t)$, given here in an obvious notation by
\begin{equation}
\epsilon_N(t)=\bigg\{\frac{1}{N}\sum_{i=1}^N \left[C_{\mathrm{numerical}}(i\Delta x,t)
-C_{\mathrm{analytical}}(i\Delta x,t)\right]^2\bigg\}^{1/2}.
\label{eq:myerr}
\end{equation}
Here, the dependency of the error on the number of grid points is indicated by the subscript $N$.  We examine this dependency by taking $t=T$ and investigating the functional relationship between $\epsilon_N(T)$ and $N$ in Figure~\ref{fig:1d_hyperdiffusion_convergence}.  The error decreases as $\epsilon_N(T)\sim N^{-2}$, consistent with the fact that that our chosen spatial discretisation of the fourth-order derivative in the hyperdiffusion equation is $O(\Delta x^2)$ (i.e., $O(N^{-2}$)).

\subsection{Implementation on GPU}
\label{sec:implementation}
In order to solve the above scheme in batched form on a GPU we follow the methodology of cuThomasBatch \cite{cuThomasBatch} with some modifications. We retain the key aspect of interleaved data layout, this means that the first row of the batch data will contain the first entry in each linear system $\mxA\vecx_i=\vecf_i$ (the subscript $i$ labels the different systems in the batch), the second row the second entry and so on. The scheme is then implemented as in the serial case, but with one thread per system. This allows the GPU threads to access the global memory with coalesced memory accesses and prevents the need to worry about the physical size limits of shared memory. Where our implementation differs, apart from the change in type of matrix, is the splitting of the initial factorisation steps from the solve steps. This allows a user who wishes to use a constant matrix repeatedly to avoid factorising at every function call, and, we will show, a user who requires a new matrix at every call is not unjustly penalised versus using the existing gpsvInterleavedBatch. Indeed in many cases we see an improvement in performance even when refactoring the matrix at every time step. 
Finally, we use the library cuSten~\cite{cuSten} to generate the right-hand side of each linear system in the batch.

The gpsvInterleavedBatch function relies on QR factorisation to solve the system of equations with householder reflection \cite{matComp}. It also relies on an interleaved data layout, thus making global data access performance identical to that of cuPentBatch. While QR factorisation is numerically stable a priori when compared to cuPentBatch it requires a greater number of operations. We note this as a flaw in cuPentBatch but we will show that for systems where the stability of the inversion is not a concern, such as in our example problem discussed in the following section where the matrix is \textbf{symmetric positive definite}, that cuPentBatch is a more efficient and faster algorithm. \textbf{It should be noted that diagonally dominant is also a valid criterion for stability when solving with cuPentBatch.}

\begin{remark}
The function gpsvInterleavedBatch uses dense QR factorisation with a zero fill pattern to accommodate the 5 diagonals while cuPentBatch is an LU factorisation without pivoting for 5 diagonals. Thus gpsvInterleavedBatch has a higher operation count than cuPentBatch. The performance benefit of this reduction is shown in Section~\ref{sec:paraversus}.
\end{remark}

\section{Performance Analysis}

For the purpose of performance analysis, we  solve a benchmark problem comprising  a series of identical one-dimensional hyperdiffusion simulations, as outlined in Section~\ref{sec:model}. To fix the emphasis on the performance analysis, each system in the batch has the same initial conditions and parameters.
Furthermore, we run each simulation for 250 time steps to average out any small variations in execution time by the computer due to scheduling, OS overhead etc. The measured time also omits any start up costs, setting initial conditions etc. 

The calculations are performed on an NVIDIA Titan X Pascal with 12GB of GDDR5 global memory and an Intel i7-6850K with 6 hyper-threaded cores. The system is running Ubuntu 16.04 LTS, CUDA v9.2.88, gcc 5.4 and has 128GB of RAM. Compiler flags used were $-O3$ $-lineinfo$ $--cudart=static$ $-arch=compute\_61$ $-code=compute\_61$ $-std=c++11$ $-lcusparse$ $-lcublas$. Also it should be noted that these benchmarks are for 64 bit doubles, so the cusparseDgpsvInterleavedBatch is the variety of the cuSPARSE function used, this choice was made as when solving numerical PDEs higher floating point accuracy is generally desirable. 

In benchmarking we have measured the following three quantities:
\begin{enumerate} 
\item The time it takes to solve a batch of hyperdiffusion equations using gpsvInterleavedBatch. We shall refer to this method as simply gpsv from now on.
\item The time it takes to solve a batch of hyperdiffusion equation using cuPentBatch, factorising the matrix once at the beginning and repeatedly solving. We shall refer to this method as cuPentBatchConstant from now on.
\item The time it takes to solve a batch of hyperdiffusion equations using cuPentBatch, resetting and factorising the matrix repeatedly at every time step. This is to examine the performance in cases where the user will want to reset the matrix at every time step.   We shall refer to this method as cuPentBatchRewrite from now on.

Even in the present context of solving parabolic numerical PDEs in batch mode, it is conceivable that the matrix $\mxA$ may change at each time step -- for instance, in situations involving mesh refinement, adaptive time stepping or where the diffusion coefficient $D$ is no longer constant.
\end{enumerate} 
\label{sec:perf}
Based on these measurements, we quantify the performance of our cuPentBatch using the following speedup ratios:
\begin{subequations}
\begin{equation}
\text{Speedup}=
\frac{\text{Time taken by gpsv}}{\text{Time taken by cuPentBatchConstant}},
\end{equation}
or
\begin{equation}
\text{Speedup}=
\frac{\text{Time taken by gpsv}}{\text{Time taken by cuPentBatchRewrite}},
\end{equation}%
\end{subequations}
depending on the context.  Hence, if $\text{Speedup}>1$, our in-house methods
are outperforming the standard gpsv.


\begin{figure}[H]
	\centering
		\includegraphics[width=0.7\textwidth]{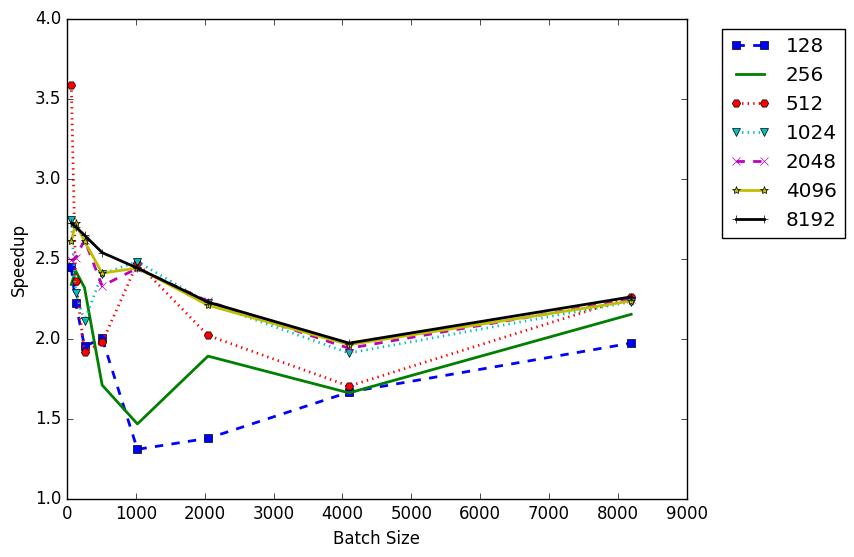}
		\caption{Speedup of cuPentBatchConstant versus gpsv. The number of unknowns for each is shown in the legend.}
	\label{fig:fixConstantNFULL}
\end{figure} 

\begin{figure}[H]
	\centering
		\includegraphics[width=0.7\textwidth]{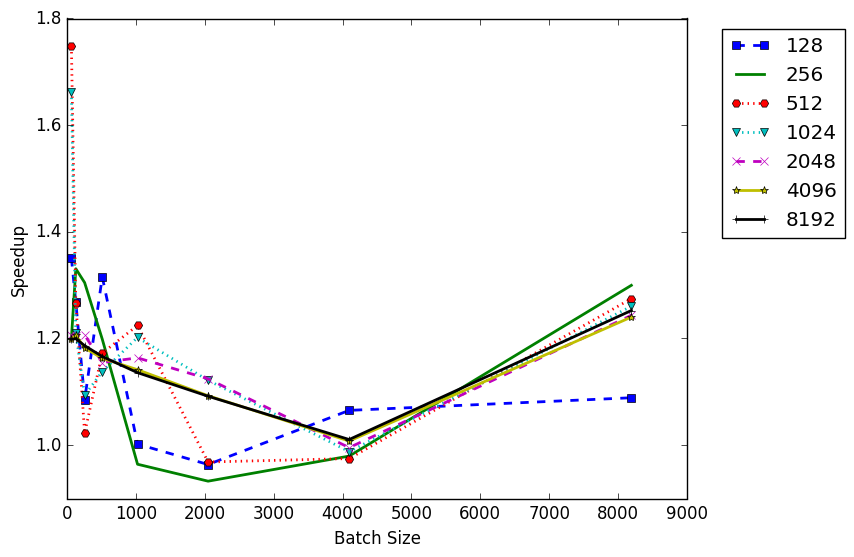}
		\caption{Speedup of cuPentBatchRewrite versus gpsv. The number of unknowns for each is shown in the legend.}
	\label{fig:fixRewriteNFULL}
\end{figure} 

\subsection{cuPentBatch vs. gpsvInterleavedBatch}
\label{sec:paraversus}
We begin by fixing the number of unknowns and varying the batch size. In Figure~\ref{fig:fixConstantNFULL} we can see clear speedup for all cases of cuPentBatchConstant, generally over $2\times$ better performance for batches with high numbers of unknowns.  Here, we see the clear advantage of the single factorisation and multiple solve over the multiple rewrites and factorisations that are required by gpsv. These batches are also small enough that they easily fit on the GPU memory thus the benchmark is free of any memory transfer penalties, only the run time of the algorithms is being compared.

In Figure~\ref{fig:fixRewriteNFULL} we can see the speedup of cuPentBatchRewrite versus the gpsv algorithm. As the matrix is now being treated as non-constant between time steps the performance is much closer to that of gpsv. Nevertheless, due to the reduced number of operations required by cuPentBatch compared to gpsv, an increase in performance can be seen.  This is most visible at larger batch numbers where there is an increase in performance such that $\text{Speedup}=1.2 - 1.3$.  

\begin{figure}[H]
	\centering
		\includegraphics[width=0.7\textwidth]{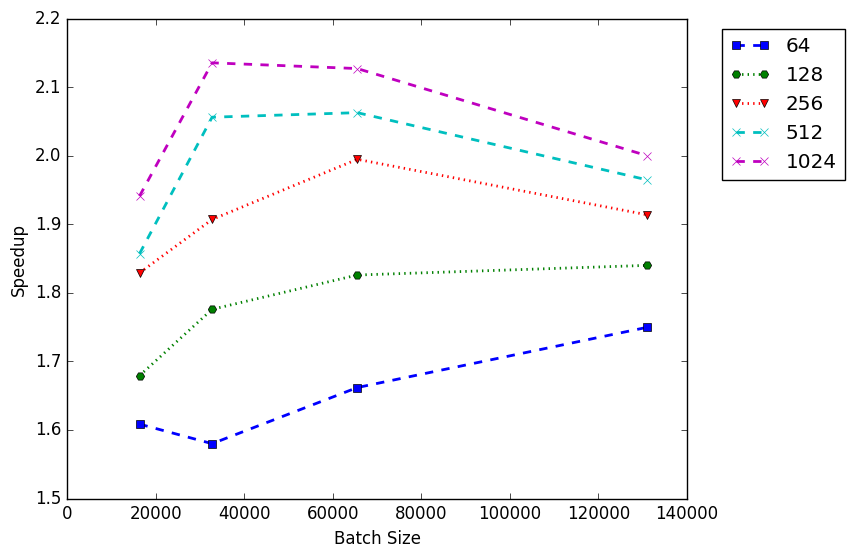}
		\caption{Speedup of cuPentBatchConstant versus gpsv for larger batch sizes $O(10^4-10^5)$. The number of unknowns for each is shown in the legend.}
	\label{fig:fixLNEQConstantNFULL}
\end{figure} 

\begin{figure}[H]
	\centering
		\includegraphics[width=0.7\textwidth]{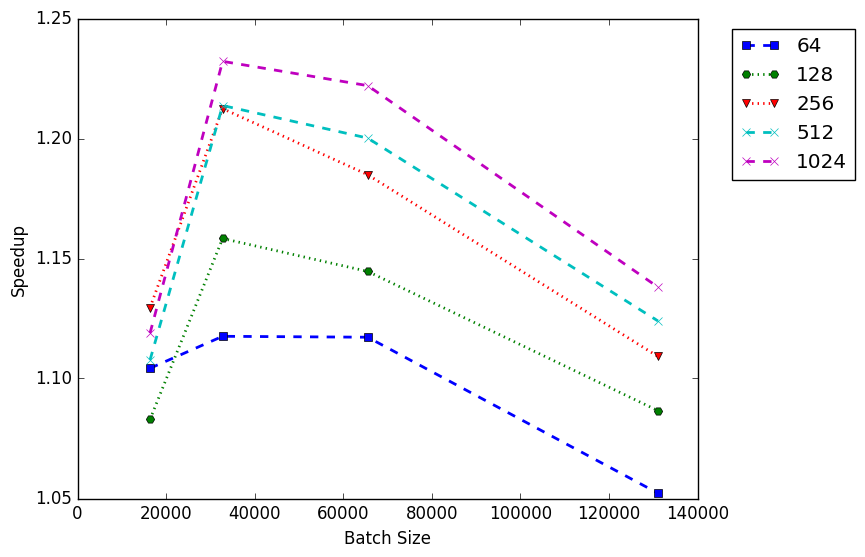}
		\caption{Speedup of cuPentBatchRewrite versus gpsv for larger batch sizes $O(10^4-10^5)$. The number of unknowns for each is shown in the legend.}
	\label{fig:fixLNEQRewriteNFULL}
\end{figure} 

Taking the batch size to an extreme $O(10^4-10^5)$ we can see the performance comparisons in Figures~\ref{fig:fixLNEQConstantNFULL} and~\ref{fig:fixLNEQRewriteNFULL}. In both we can see that the improvement drops away as batch size increases but cuPentBatch is still faster in both cases, particularly for the higher unknown sizes of 512 and 1024. Thus it is clear for almost all batch sizes cuPentBatch is the better performer regardless of fixing a constant matrix or using a new one for every time step. 

\begin{figure}[H]
	\centering
		\includegraphics[width=0.7\textwidth]{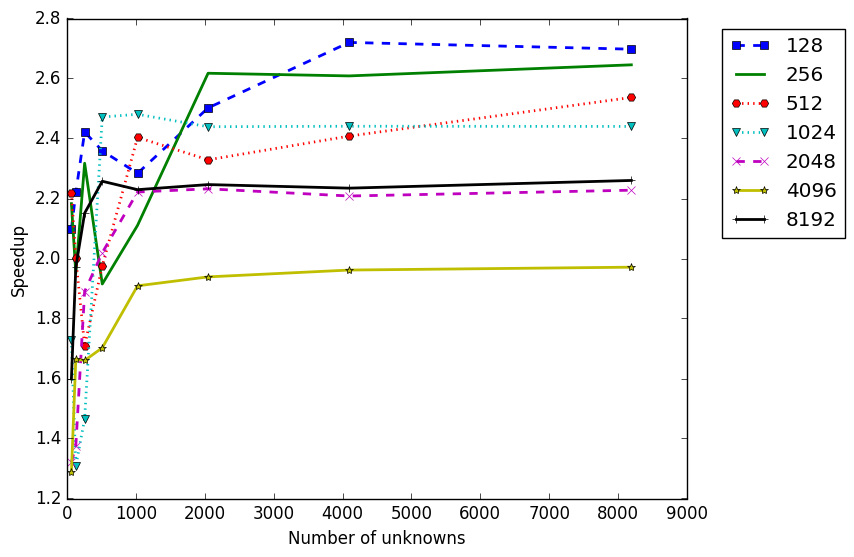}
		\caption{Speedup of cuPentBatchConstant versus gpsv. The batch size for each is shown in the legend.}
	\label{fig:fixConstantNEQ}
\end{figure} 

\begin{figure}[H]
	\centering
		\includegraphics[width=0.7\textwidth]{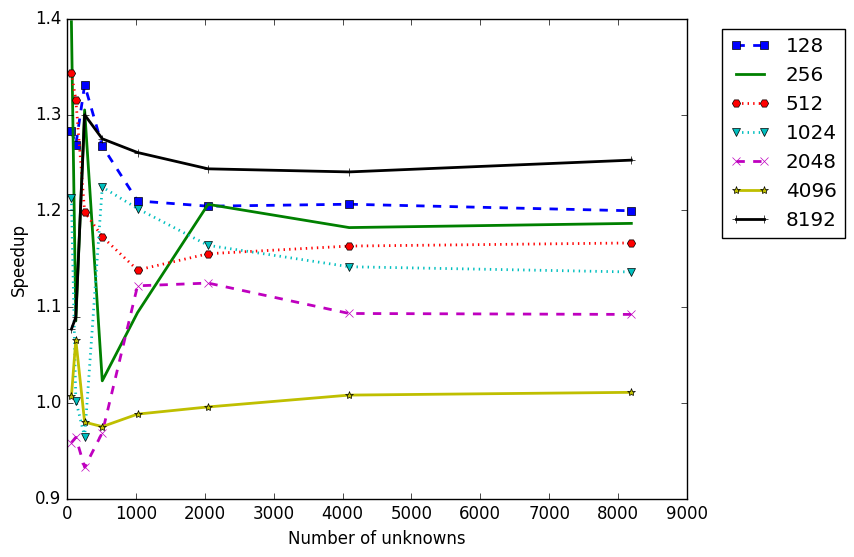}
		\caption{Speedup of cuPentBatchRewrite versus gpsv. The batch size for each is shown in the legend.}
	\label{fig:fixRewriteNEQ}
\end{figure} 

\begin{figure}[H]
	\centering
		\includegraphics[width=0.7\textwidth]{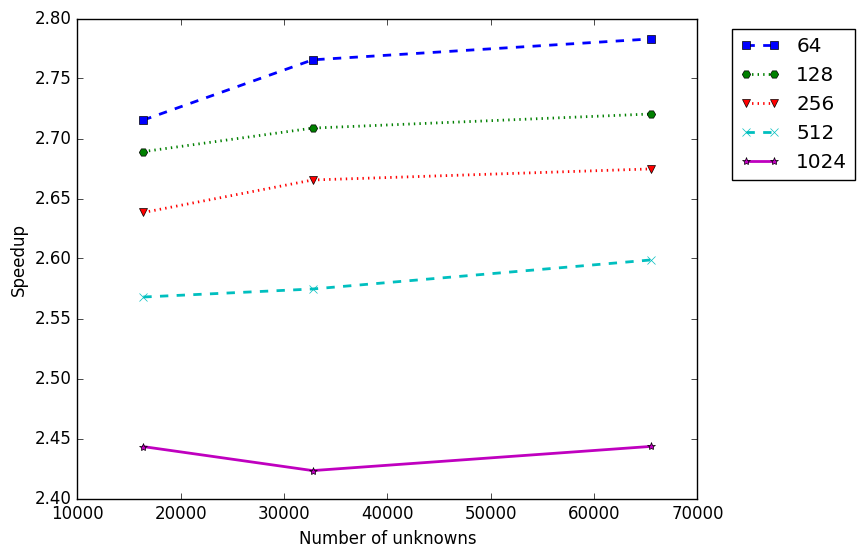}
		\caption{Speedup of cuPentBatchConstant versus gpsv  for large numbers of unknowns $O(10^4)$. The batch size for each is shown in the legend.}
	\label{fig:fixLNFULLConstantNEQ}
\end{figure}

\begin{figure}[H]
	\centering
		\includegraphics[width=0.7\textwidth]{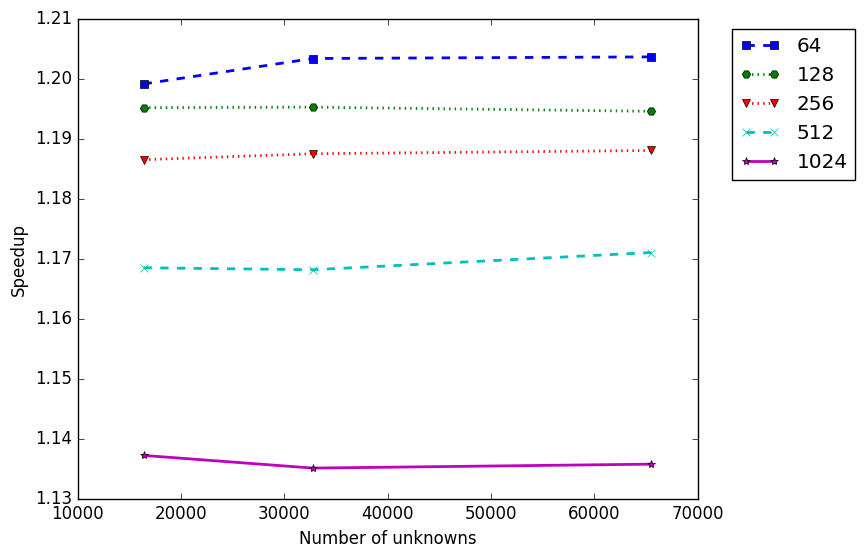}
		\caption{Speedup of cuPentBatchRewrite versus gpsv for large numbers of unknowns $O(10^4)$. The batch size for each is shown in the legend.}
	\label{fig:fixLNFULLRewriteNEQ}
\end{figure}

We now perform a further analysis where we keep the size of the batch fixed and vary the number of unknowns. For cuPentBatchConstant in Figure~\ref{fig:fixConstantNEQ} we again see significant speedup, especially at higher numbers of unknowns where speed up is well over $2\times$. Similarly in Figure~\ref{fig:fixRewriteNEQ} we see better performance.   It is clear that at high numbers of unknowns cuPentBatch performs significantly better than gpsv. This result is further confirmed in Figures~\ref{fig:fixLNFULLConstantNEQ} and~\ref{fig:fixLNFULLRewriteNEQ} where the resolution each system highly resolved with a moderate batch size. Summarising,  cuPentBatch outperforms gpsvDInterleavedBatch in terms of scaling the number of unknowns in a system.

\subsection{cuPentBatch vs. Serial}
Given that we have established the speedup available to us over gpsvDInterleavedBatch we now show that for solving batches of pentadiagonal systems cuPentBatch is far superior to doing the same calculation in serial. The serial benchmark was run on the same machine as the cuPentBatch with similar compiler optimisations turned on. The data is laid out in a standard format, not interleaved. This is to allow the memory to be accessed in C's preferred row major format, one hyperdiffusion system per row. 

\begin{figure}[H]
	\centering
		\includegraphics[width=0.7\textwidth]{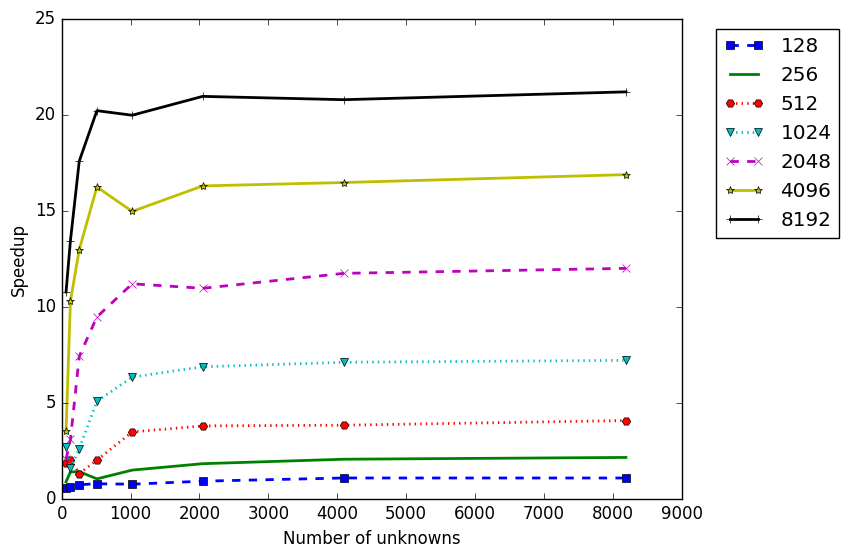}
		\caption{Speedup of cuPentBatchConstant versus serial. The batch size for each is shown in the legend.}
	\label{fig:serialConstantNEQ}
\end{figure} 

\begin{figure}[H]
	\centering
		\includegraphics[width=0.7\textwidth]{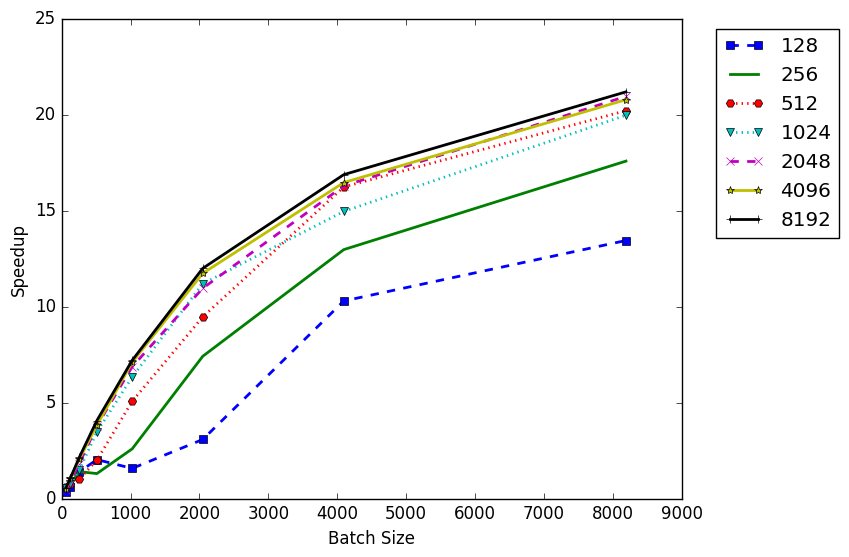}
		\caption{Speedup of cuPentBatchConstant versus serial. The number of unknowns for each is shown in the legend.}
	\label{fig:serialConstantNFULL}
\end{figure} 

\begin{figure}[H]
	\centering
		\includegraphics[width=0.7\textwidth]{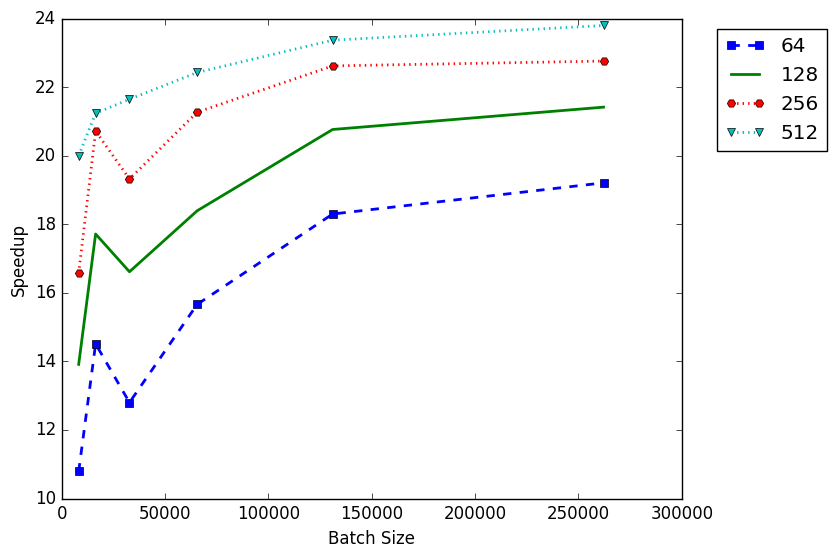}
		\caption{Speedup of cuPentBatchConstant versus serial with large batch size. The number of unknowns for each is shown in the legend.}
	\label{fig:serialExtConstantNFULL}
\end{figure} 

In Figure~\ref{fig:serialConstantNEQ} we see a speedup comparison of cuPentBatchConstant and the serial version of the code keeping the batch size constant and varying the number of unknowns in a system. Again we take the time taken to execute the serial code and divide this by the time taken by the GPU code. At low batch numbers the speedup is minimal as the serial aspect of the pentadiagonal inversion dominates. As the size of the batch increases so does the speed up, with over $20\times$ faster performance for systems with a batch size of 8192. This finding is reinforced in Figure~\ref{fig:serialConstantNFULL} where the number of unknowns is kept constant and the batch size is varied. Significant speedup can only be seen at higher batch sizes with over $10\times$ faster for most systems with a batch size $>2048$. Taking the batch number to an extreme in Figure~\ref{fig:serialExtConstantNFULL} we can further see how cuPentBatch scales well in terms of increasing batch size.

We see the clear presence of Amdahl's Law in these graphs, particularly in Figure~\ref{fig:serialConstantNEQ}. We see the benefits of parallelising the code until the serial aspect of the pentadiagonal solve begins to dominate. The performance increases then level off at this point and no increased speedup can be obtained from the system. Amdahl's Law can also be seen in Figure~\ref{fig:serialExtConstantNFULL} as the performance increases for solving high batch numbers are ultimately bound by the number of unknowns $N$.  

\subsection{cuPentBatch vs. OpenMP}
\begin{figure}[H]
	\centering
		\includegraphics[width=0.7\textwidth]{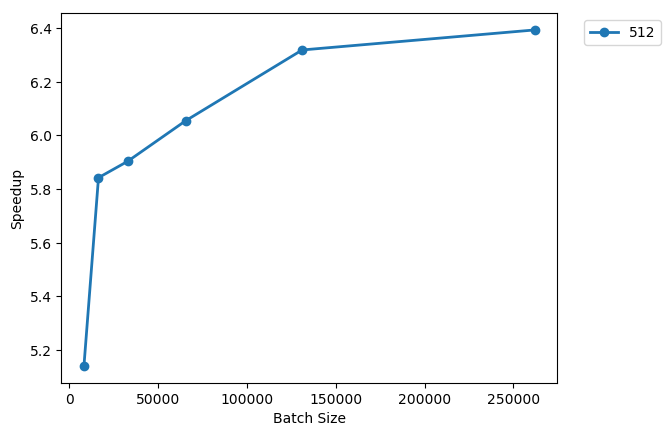}
		\caption{Performance of cuPentBatch versus an OpenMP version of the serial code, the number of unknowns was fixed at 512 and 8 threads were used for the OpenMP benchmark.}
	\label{fig:ompCompare}
\end{figure}

The OpenMP implementation is the same as the serial code except we have parallelised the loop over the batches,  the speedup is measured as the time taken for the OpenMP version divided by cuPentBatchConstant. For the OpenMP benchmark we ran the batch solver with 512 unknowns and varied with high batch numbers, we choose this method for comparison as we are most concerned with scaling at high batch numbers as such problems benefit most from parallelisation. For consistency the same system was used as for the previous computations. The number of threads was set at 8 as this was the highest power of 2 available. The results in Figure~\ref{fig:ompCompare} show a speedup of $5\times$ to $6\times$ in every case, thus demonstrating a substantial improvement in performance.  


\section{Discussion and Conclusions}
\label{sec:conc}

Summarising, we have introduced a new pentadiagonal solver (cuPentBatch) for implementation on NVIDIA GPUs, specifically aimed at solving large numbers of pentadiagonal problems in parallel, in batch mode.
We have shown that our own solver cuPentBatch is superior in terms of performance and efficiency to that of the standard existing NVIDIA pentadiagonal solver, gpsvInterleavedBatch. 
Our method further exhibits substantial performance speedup when compared with serial and OpenMP implementations.
Our method is particularly useful for solving parabolic numerical PDEs, where the matrix to be solved at each time step is constant and \textbf{symmetric positive definite}.
We have demonstrated a potential application of our method in the context of parameter studies, whereby the pertinent PDE possesses parameters which may be varied over different simulations to produce different solution types. By solving multiple instances of the PDE in batch mode, our method can speed up such parameter studies.

A further application of our method may in future be found in solving parabolic numerical PDEs in two and three dimensions -- here the pertinent parabolic PDE is typically solved using using an implicit temporal discretisation and a standard finite-difference spatial discretisation.  The resulting matrix to be inverted at each time step can be reduced to a series of one-dimensional problems using the alternating-direction-implicit (ADI) technique~\cite{douglas1955numerical}.  In this scenario, the present pentadiagonal batch solver may prove useful for parallelising this wide variety of  numerical algorithms.


\subsection*{Acknowledgements}
Andrew Gloster acknowledges funding received from the UCD Research Demonstratorship.   Khang Ee Pang acknowledges funding received from the UCD School of Mathematics and Statistics the Summer Research Projects 2018 programme.
All authors gratefully acknowledge the support of NVIDIA Corporation with the donation of the Titan X Pascal GPUs used for this research.  The authors also thank 
Lung Sheng Chien and Harun Bayraktar of NVIDIA for helpful discussions throughout the project.

\end{document}